Impact of the various spin and orbital ordering processes on multiferroic properties of orthovanadate $DyVO_3$

Q. Zhang[1,4*], K. Singh[1,5], C. Simon[2], L. D. Tung[3], G. Balakrishnan[3] and V. Hardy[1]

[1]Laboratoire CRISMAT, UMR 6508, CNRS ENSICAEN, 6 Boulevard du Maréchal Juin, F-14052 Caen Cedex 4, France

[2]Institut Laue Langevin, 6 rue Jules Horowitz BP 156 F-38042 Grenoble Cedex 9, France

[3]Department of Physics, University of Warwick, Coventry CV4 7AL, United Kingdom

[4]Ames Laboratory and Department of Physics and Astronomy, Iowa State University, Ames, Iowa 50011, United States

[5]UGC-DAE Consortium for Scientific Research, University Campus, Khandwa Road, Indore 452001, India

Abstract

The orthovanadate $DyVO_3$ crystal, known to exhibit multiple structural, spin and orbital ordering transitions, is presently investigated on the basis of magnetization, heat capacity, resistivity, dielectric and polarization measurements. Our main result is experimental evidence for the existence of multiferroicity below a high $T_C$ of 108 K over a wide temperature range including different spin-orbital ordered states. The onset of ferroelectricity is found to coincide with the antiferromagnetic C-type spin ordering transition taking place at 108 K, which indicates that $DyVO_3$ belongs to type II multiferroics exhibiting a coupling between magnetism and ferroelectricity. Some anomalies detected on the temperature dependence of electric polarization are discussed with respect to the nature of the spin-orbital ordered states of the V sublattice and the degree of spin alignment in the Dy sublattice. The orthovanadates $R$VO$_3$ ($R$ = rare earth or Y) form an important new category for searching for high-$T_C$ multiferroics.

## I. INTRODUCTION

Multiferroic materials, with coexistence of magnetic ordering and ferroelectricity, have recently attracted immense attention, aimed both at studying the fundamental properties and at exploiting them in multifunctional device applications[1-3]. In conventional type-I multiferroics[4,5], the origins of the ferroelectric and the magnetic order are not related, which leads to weak magnetoelectric coupling. In contrast, the type-II multiferroics (related to the so-called spin driven ferroelectricity)[6,7] usually exhibit a lower electric polarization, but the ferroelectricity is directly coupled to the magnetic order and therefore a strong intrinsic magnetoelectric coupling can be induced. To date, spin driven ferroelectricity has been discovered in various systems, such as boracite $Ni_3B_7O_{13}Br$[8], chromium chrysoberyl $Cr_2BeO_4$[9], perovskite manganites $R$MnO$_3$ ($R$=Tb, Dy, Ho, Tm, Y)[7], $R$Mn$_2$O$_5$ ($R$=Tm, Ho, Dy, Tb, Y, Er)[10,11,12] $Ni_3V_2O_8$[13], $FeVO_4$[14], $MnWO_4$[15], $CoCr_2O_4$[16], $FeCr_2O_4$[17], $FeV_2O_4$[18], $CuCrO_2$[19], $CuBr_2$[20], $R$CrO$_3$($R$= Sm, Gd, Tb, Er,Tm)[21], $CuO$[22], $CaMn_7O_{12}$[23], and $SmFeO_3$[25,26] etc[6,7]. It is known that the leakage is a major problem for measurement of ferroelectricity and for applications of high-temperature multiferroics. In this regard, it is beneficial if a multiferroic materials could exhibit both a high ferroelectric critical temperatures ($T_C$) and a very low dielectric loss (<<0.1[18]). On the other hand, if spin driven ferroelectricity would occur in materials that exhibit several magnetic or orbital ordering transitions, multiple anomalies in temperature dependence of electric polarization are expected to take place in different temperature windows. Such materials could thus provide good candidates to allow investigating the effect of different magnetic or orbital orderings on ferroelectricity in the same single-phase material. Therefore, it is of interest to search for novel magnetic multiferroics with high ferroelectric critical temperatures, very low dielectric loss, and successive/multiple anomalies in the ferroelectricity over a wide temperature region.

Compared to manganites with orbital-active $e_g$ electrons, the perovskite-type vanadium oxides $R$VO$_3$ ($R$=rare earth elements and Y) with orbital-active $t_{2g}$ electrons is another type of correlated electron system and may be expected to undergo a weaker Jahn-Teller interaction.[27-29] The relationship between spin, orbital and lattice degrees of freedom in such systems is more subtle and complex than that observed in manganites. The $R$VO$_3$ compounds have a $Pbnm$ orthorhombic lattice at room temperature and exhibit two distinct spin-ordering (SO) and orbital-ordering (OO) states depending on the $R^{3+}$ ionic radius and/or temperature.[28, 30-33] Interestingly, switching between these two spin/orbital ordering states can be induced by external stimuli like a magnetic field, when $R$ is Dy or Ho.[28] All members of the $R$VO$_3$ family exhibit a structural transition at $T_{OO1}$ (ranging from 150 to 200 K) from $Pbnm$ orthorhombic to $P2_1/b$ monoclinic with G-type OO (see Fig. 1 (a)). In this G-type OO, one electron–occupies the $d_{xy}$ orbital, while another electron occupies the $d_{yz}$ or $d_{zx}$ orbitals that are staggered in all three spatial directions.[28, 30] With decreasing temperature, all $R$VO$_3$ compounds exhibit a magnetic transition [28, 31] at $T_{SO1}$, from the paramagnetic state to C-type SO. This ordering of the V$^{3+}$ spins corresponds to an antiferromagnetic (AFM) alignment within the $xy$ plane, together with a ferromagnetic alignment along the $z$ axis, as shown in Fig. 1 (a). The AFM Néel temperature (named $T_{SO1}$ here) monotonously increases from 90 K (Lu) to 140 K (La) because of the enhancement of the spin and orbital exchange interactions between the nearest neighbor V sites induced by decreasing V-O-V bond angle distortion with increasing ionic radius.[28] It turns out that DyVO$_3$ is located at the vicinity of the boundary (1.09 Å) of the $R^{3+}$-ion radius[34] and also at the phase boundary between the two competing spin-orbital-ordered states[34], a peculiarity which makes its structural and physical properties very complex below $T_{SO1}$. On the basis of previous studies[28, 30, 33-37], the sequence of the transitions affecting the spin, orbital and lattice degrees of freedom in DyVO$_3$ is summarized in Fig. 1 (c). The G-type OO state in DyVO$_3$ evolves to C-type OO below $T_{SO2}$ (55 -62 K) [28, 31, 36] with the alternate $d_{xy}^1 d_{yz}^1 / d_{xy}^1 d_{zx}^1$ electron configuration in the *xy* plane and the same one along the *z* axis and the structure changes back to the *Pbnm* orthorhombic structure again. Simultaneously, the C-type SO reorients into G-type SO at $T_{SO2}$ with an AFM coupling between the V$^{3+}$ moments (//c) in all three directions (see Fig. 1 (b)). In addition, DyVO$_3$[28,35,37] exhibits an additional reentrant transition to C-type SO/G-type OO and a structural transition from *Pbnm* orthorhombic to *P2$_1$/b* monoclinic as the temperature is lowered below $T^*_{SO2}$ (18-22 K). It should be noted that recent X-ray and Raman scattering experiments[28] have shown that the volume fraction of the C-type SO/G-type OO state below $T^*_{SO2}$ is only about 1/3 and that the rest remains in the G-type SO/C-type OO state, resulting in a phase coexistence state below $T^*_{SO2}$. Motivated by the particular richness of the spin and orbital ordering processes encountered in DyVO$_3$ crystal, we have selected this system to investigate the possible occurrence of multiferroic behavior in the *R*VO$_3$ family.

## II. EXPERIMENTAL DETAILS

The DyVO$_3$ crystals have been grown with the floating-zone method using a high-temperature xenon arc furnace. The details of the crystal growth are similar to those reported previously for other *R*VO$_3$ crystals.[32, 38] The crystalline orientation was confirmed by using back-reflection x-ray Laue photographs. The temperature/field dependence of magnetization for H//b and resistivity were recorded by means of a Physical Properties Measurement System (PPMS, Quantum Design). Heat capacity measurements were carried out in the same PPMS using a semiadiabatic relaxation method.

The complex dielectric permittivity was measured on a parallel plate capacitor geometry and silver-paste electrodes were made on two different capacitors to apply an electric field along the *c* or the *b* axis and the magnetic field is applied along the b axis. The dielectric permittivity ε' and the losses tanδ were measured using an Agilent 4284A LCR meter during both cooling and warming processes (0.5 K/min) with a 0.5 V *ac* bias field at different frequencies (5 kHz to 100 kHz). The electric polarization with E//b was measured with a Keithley 6517 A electrometer in a coulomb mode with automatic current integration facility.

To align the ferroelectric domains, a static electric poling field of +400 kV/m was applied along the *b* or the c axis during cooling from 120 K down to 8 K. After this poling field was removed at 8 K, we did short circuit and P was recorded as a function of time for 6400 seconds at 8 K to reach a stable state and to remove possible stray or trapped charges. The detailed experimental procedure and related experimental data can be found in the "Supplemental Material"[39]. The temperature dependence of P was subsequently recorded during warming (0.5 K/min) in zero electric field. After this, the direction of static electric poling field (-400 kV/m) was changed and the same experimental procedure was carried out to measure the temperature dependence of the electric polarization with E//b. As known, the polarizations in the type-II multiferroics are usually typically 3-4 orders of magnitude lower traditional ferroelectricity, like $10^2$ uC/cm$^2$ in BiFeO$_3$,[40] and the coercive electric fields in such type-II multiferroics are usually very high, which makes the P(E) hysteresis loop measurements hardly achieved. Thus, the above experimental procedure and the symmetric reversal of P upon reversing the poling electric field are widely used ways to check the presence of ferroelectric polarization, as has been adopted in many previous reports.[17-18,21-22,40-41] We have also measured the resistivity and dielectric loss and paid cautious attention to exclude other possible contribution, such as leakage contribution or trapped charges, as discussed in the text.

## III. RESULTS AND DISCUSSION

### A. Magnetic properties

The temperature dependence of the field-cooled-cooling (FCC) and field-cooled-warming (FCW) magnetization of $DyVO_3$ in a magnetic field of 20 Oe parallel to the $b$ axis is shown in Fig. 2 (a). There is an increase in magnetization at $T_{SO1} \sim 108$ K for both the FCC and the FCW process, corresponding to a magnetic transition from paramagnetic (PM) to AFM C-type SO transition.[28, 34, 37] The increase of magnetization at $T_{SO1}$ has been reported to reflect the spin canting of V moments related to the antisymmetric Dzyaloshinskii-Moriya (DM) interaction[34, 37]. Note that for the $RVO_3$ compounds with small ionic radius such as Dy, Y, Ho and Er,[37-43] a weak $z$-component of the G-type exists along with the previously reported strong C-modes in the $xy$-plane.[30,34,37-43] Thus, the magnetic structure below $T_{SO1}$ in these systems is not purely collinear C-type SO. Instead, the AFM-aligned V $3d$ spins are a superposition of C-modes in the $xy$ components (AFM alignment in the $xy$ plane and ferromagnetic alignment along the $z$ axis) and the weakest G-type order of the $z$ component, forming noncollinear vanadium spins with $CxCyGz$ mode in $DyVO_3$[30,34,37-43]. Upon decreasing the temperature to $T_{SO2}$, a rapid decrease in the magnetization along *b* axis takes place for both the FCC and the FCW process. This is ascribed to the reorientation[28,35] from C-type SO to collinear G-type SO with the V spins parallel to the *z* axis, the latter configuration being not expected to favor large spin canting. There is a significant thermal hysteresis around $T_{SO2}$ between the FCC and the FCW magnetization curves, indicating that this transition is first order. The transition temperature is found to be 55.2 K and 62.3 K in the cooling and warming process, respectively. The thermal hysteresis of 7 K is in line with the results (57 and 64 K) reported by Miyasaka et al.[35] For the sake of simplicity, we will only use the value (~62 K) for the warming process in the rest of the paper. Around $T^*_{SO2}$ (~18 K), the magnetization increases rapidly and there is a change in the slopes of the magnetization as well as an anomaly in the inverse susceptibility (not shown), which shows a reentrant transition from G-type SO/C-type OO to C-type SO/G-type OO[28, 35]. These magnetic characterizations in the present crystal are consistent with previous reports[24,28,35]. It is worthwhile noting that the Dy moments have been reported to become fully ordered below $T^*_{SO2}$[42]. The increase of the magnetization at $T^*_{SO2}$ is therefore due to the combined contribution from the spin canting of the V spins and the ordered Dy moments.

In the temperature region between $T^*_{SO2}$ and $T_{SO2}$, a field-induced metamagnetic transition from G-type SO/C-type OO to C-type SO/G-type OO is observed in DyVO$_3$, as can be seen in the representative isothermal magnetization curve at 30 K with H//*b* in Fig. 2 (b). For the virgin curve, the magnetization starts to increase rapidly around a threshold field *Ht* (~1.6 T) due to the metamagnetic transition[28, 34, 35, 37] from G-type SO/C-type OO to more stable C-type SO/G-type OO, a magnetic order in which spin canting allows a larger net magnetization to be reached in a magnetic field[37]. A large magnetic hysteresis is observed as the field is ramped up and down, marking again the first-order nature of the transition between G-type SO/C-type OO and C-type SO/G-type OO. Since the maximum magnetic moment of V $3d$ is only 2 $\mu_B$, the high saturation magnetization of around 8 $\mu_B$ reached above $Ht$ suggests that the field-induced spin/orbital transition occurs concomitantly with the polarization of Dy $4f$ moments by the external field, which can be explained by the coupling between the Dy $4f$ and the V $3d$ spins.[28] In addition, the high saturation magnetization for H//b suggests that the $b$ axis is an easy magnetization direction, in accordance with the *Refs.* 28 and 35, in which both *a* and *b* axes were reported to be easy axes while the *c* axis is magnetic hard axis.

### B. Heat capacity

The temperature dependence of $Cp$(T)/T below 180 K in zero magnetic field for the warming process is shown in Fig. 3. The peak at 108 K marks $T_{SO1}$, while the rapid decrease at 62 K with decreasing temperature corresponds to $T_{SO2}$. Another anomaly can be observed at 18 K (see inset of Fig. 3), which marks the reentrant transition temperature $T^*_{SO2}$. The three transition temperatures are in good agreement with the values observed in the magnetization measurements. The present heat capacity results are different from those reported in *Ref.* 30, where no anomaly is observed at $T^*_{SO2}$. We suggest that, as compared with $T_{SO2}$, the weaker anomaly in $Cp$(T) at $T^*_{SO2}$ should be ascribed to the incomplete spin/orbital transition, as mentioned above. In contrast, at $T_{SO2}$, the C-type SO/G-type OO state transforms completely to pure G-type SO/C-type OO, leading to a sharp jump in $Cp$(T)/T. Similar features are observed in the dielectric properties, as will be shown below.

### C. Resistivity, dielectric permittivity and loss

The resistivity of $DyVO_3$ is found to increase rapidly with decreasing temperature (see inset of Fig. 2 (b)) without the magnetic-field induced magnetoresistance effect. At 227 K, the resistivity is already as high as $6 \times 10^4$ $\Omega$ cm (resistance is $4.3 \times 10^5$ $\Omega$), suggesting that the resistivity of DyVO$_3$ below 120 K is far higher than this magnitude below ≈120 K. Figures 4 (a) and (b) present the temperature dependence of dielectric permittivity ε' during the warming process at 5 kHz with ac bias E//b and E//c, respectively. One weak kink with a slope change corresponding to a peak in the first derivative of the ε'(T) curve is observed around $T_{SO1}$ with E//b and E//c. When the frequency increases to be 100 kHz, the kink in ε'(T) curve evolves to one concave cusp, as shown in Fig. 5 (a) and (c). Such type of dielectric anomaly (a change of slope or a peak in dε'(T)/dT) has been reported at the ferroelectric critical temperature for many spin driven ferroelectricity, such as Ca$_3$Co$_{2-x}$Mn$_x$O$_6$ (x=0.96)[44], ACr$_2$O$_4$ (A=Fe or Cr)[17], FeV$_2$O$_4$[18], $R$CrO$_3$($R$= Sm, Gd, Tb, Er, Tm)[21]. As also shown in Fig. 5 (a) and (c), a rapid jump in ε' and a large thermal hysteresis between cooling and warming processes around $T_{SO2}$ marking the first-order transition. We point out that ε' is not superimposed mainly above $T_{SO2}$ for cooling and warming process. The exact origin of this feature is not clear, but this is probably ascribed to the influence of such first-order transition at $T_{SO2}$ after the different thermal processes. In addition, a kink is also observed at $T^*_{SO2}$ (=18 K) with E//b and E//c. Note that the locations of these three anomalies in ε' are frequency independent (not shown) and coincide well with the three transitions observed in the magnetization and heat capacity measurements.

It is worthwhile pointing out that the value of the dielectric loss in dielectric measurements reflects the insulating behavior of the sample. For a real capacitor (good insulator), the value of tan$\delta$ should be close to zero. The temperature dependence of the dielectric loss (tan$\delta$) recorded at 100 kHz is shown in Fig 5 (b) and (d). At 120 K, the value of tan$\delta$ is quite low (≈0.02) for E//b or c and further decreases to close to zero, followed by a weak increase with decreasing temperature. Combining the very low dielectric loss with the resistivity measurements (see Fig. 2 (b)), we can conclude that DyVO$_3$ is highly insulating below ≈120 K. Moreover, the locations of three anomalies in ε' are frequency-independent. Thus, we can exclude possible extrinsic factors,[20] like trapped interfacial charge carriers, and state that these measurements reflect intrinsic electric transitions at $T_{SO1}$, $T_{SO2}$, and $T^*_{SO2}$. Moreover, the anomalies at $T_{SO1}$, $T_{SO2}$ and $T^*_{SO2}$ are also found in the tan$\delta$ vs T curve, for both the warming and the cooling process (see Fig. 5 (b) and (d)). Note that the formation of electric dipoles leads to an increase in the dielectric loss at $T_{SO1}$, which can be seen more clearly under a low frequency, like 5 kHz in the inset of Fig. 5 (c) and (d). Note that although there are no clear differences in the locations of the three anomalies in the ε'(T) curve with E//b and E//c, ε' shows anisotropic behavior at $T_{SO2}$, i.e. a rapid decrease of ε' for E //c in contrast with rapid increase for E //b during cooling, exhibiting the anisotropic features of the dielectric properties. It is worthwhile pointing out that one observes another broad hump in tan$\delta$ which is centered around ~36 K for E//b, but this feature is absent for E//c, which further indicates the anisotropic character of the dielectric properties of DyVO$_3$.

To study the effect of a magnetic field on the dielectric properties, we have measured ε' *vs* T at 100 kHz in a small magnetic field of 5 kOe parallel to the *b* axis for both orientations (E//*b* and E//*c*), as shown in Figs. 6 (a) and (b). Interestingly, a clear effect of magnetic field on ε' is observed at $T_{SO1}$, i.e. a sudden H-induced dip in ε' for E //b. In addition, a small magnetic field of 5 kOe shifts the anomaly at $T_{SO2}$ slightly to lower temperatures (~ 1 K) for E//*b*. No such effects are observed for E//*c*, which implies the possible anisotropic effect of magnetic field on ε'.

### D. Electric polarization with multiple anomalies

To investigate the presence of ferroelectricity, we have measured temperature dependence of electric polarization (P) with an electrometer directly in the Coulombic mode with E parallel to *b* and *c* axis. The detailed measurement procedure and related experimental data can be found in the "experimental details" part and "SUPPLEMENTARY INFORMATION". As shown in Fig. 7, one observes the appearance of polarization in $DyVO_3$ below the magnetic transition at $T_{SO1}$ for E//b or E//c, a feature pointing to the magnetic origin of the ferroelectricity in this system. With decreasing temperature, two additional anomalies can be detected in $P_b(T)$ curve around $T_{SO2}$ and around 36 K, corresponding to the two peaks in the first derivative of $P_b$ ($dP_b/dT$) (see inset of Fig. 7 (a)). One can also observe an anomaly in $P_c(T)$ around $T_{SO2}$. However, there is no clear anomaly in $P_c(T)$ around 36 K and an additional decrease in the electric polarization is only observed below $T^*_{SO2}$ in $P_c(T)$ as shown in Fig. 7 (b). This overall behavior demonstrates a close interplay between the nature of the spin/orbital ordering and the ferroelectric properties. The remnant polarization values along the *b* and *c* axes are 320 and 700 $\mu C/m^2$, respectively.

Much cautious attention has been paid to exclude any artifact effect and confirm the ferroelectricity. The electrical current *i* between the two electrodes attached on a parallel plate can be given as[45]

$$i = C\frac{dV}{dt} + A\frac{dP}{dt} + \frac{V}{R} \qquad (1)$$

where *V* is applied voltage, $C = \varepsilon\varepsilon_0 A/L$ is the capacitance, *A* and *L* are contact area and *L* is the thickness of the crystal, *P* is the ferroelectric polarization, and *R* is the sample resistance. The three terms in Eq. 1 from left to right are capacitive, ferroelectric and resistive currents. The high resistance without the magnetoresistance effect, the very low dielectric loss, and frequency-independent anomalies in the dielectric permittivity indicate the high-insulating behavior below ≈120 K, which excludes significant resistive current and also the influence of Maxwell-Wagner effect in DyVO$_3$. Moreover, the current and voltage phase angle derived from the very low dielectric loss in the dielectric measurements is around -89°, corresponding to the ideal value of a perfect capacitor. In addition, we have done short circuit and measured P as a function of time for long time (6400 seconds) to remove stray or trapped charges (if any) and to reach a stable state prior to the P *vs* T measurements. This procedure safely excludes a possible contribution from leakage contribution or trapped charges to pyroelectric current.[17,18,40,41,45,46] The P *vs* T measurements have been conducted at zero bias voltage with increasing temperature to exclude the capacitive current and any artifact resulting from the variation of the electric field.[45-46] It must also be emphasized that reversing the poling electric field leads to a symmetric negative polarization indicative of switchable polarization (see Fig. 7 (a)), which further confirms[17,18,21-22,41] the electric polarization below $T_{SO1}$(≈108 K) in DyVO$_3$. Note that the ferroelectricity in DyVO$_3$ takes place below a high temperature of 108 K (well above liquid nitrogen temperature) and exhibits substantial remnant polarization reaching ~ 700 μC/m$^2$ for E//c (with E=400 kV/m). The saturation polarization is expected to be higher at high poling electric field.

### E. Origin of the ferroelectricity in $T_{SO2}$<T<$T_{SO1}$ and 36 K<T<$T_{SO1}$

Below $T_{OO1}$ (~195 K), the G-type OO is formed with the orthorhombic-monoclinic structural transition, which however does not induce the electric polarization. The electric polarization only emerges below $T_{SO1}$, where the G-type OO and fundamental structure are unchanged and only the spin ordering changed from paramagnetic to long-range ordered C-type SO, which indicates that the presence of ferroelectricity below $T_{SO1}$ is mainly driven by the spin ordering and therefore,

DyVO$_3$ belongs to type II multiferroics. As mentioned above, the V$^{3+}$ magnetic moments exhibit noncollinear ordering with the modes *CxCyGz* [37-43] in the temperature region $T_{SO2}$ < T < $T_{SO1}$. It has been established that the origin of polarization in the noncollinear magnetic structure can be explained by the $S_i \times S_j$ type spin-current model or the antisymmetric DM interaction[6, 18, 47]. Such model has successfully explained the origin of polarization in many noncollinear magnetic systems[10,13-18,47,48]. Recently, Kaplan[49] has further developed the conventional spin-current model by adding additional terms which were omitted previously in the existing microscopic models. It has been demonstrated that a pair of magnetic atoms with canted spins *S$_a$* and *S$_b$* can give rise to an relatively large electric dipole moment *P*. Thus, in DyVO$_3$, based on the (extended) spin-current model, the noncollinear V 3*d* spins in DyVO$_3$ with weak spin canting due to the existing antisymmetric DM exchange interaction[37] imposed to the monoclinic *P2$_1$/b* space group probably contribute to some extent to the polarization along *b* and *c* axes in the temperature interval $T_{SO2}$ < *T* < $T_{SO1}$.

However, the case here is not so simple. In the temperature region of 36 *K* < *T* < $T_{SO2}$, the noncollinear C-type SO reorients into collinear G-type SO and simultaneously, the G-type OO switches to C-type OO. The (extended) spin-current model cannot predict the polarization due to the collinear V$^{3+}$ magnetic structure in the purely G-type SO. It should be noted that the magnetic structures change from C-type SO to G-type SO in DyVO$_3$ at $T_{SO2}$, but the direction of the electric polarization does not change despite the different spin orientation (see Fig. 7), which is an indication[59] that the $S_i \bullet S_j$-type exchange striction mechanism plays an important role in the polarization throughout $T_{SO2}$, i.e., not only below $T_{SO2}$, but also above $T_{SO2}$ ($T_{SO2}$<T<$T_{SO1}$). This is in contrast with the spin-current model, based on which the direction of P is strongly dependent on the direction of $S_i \times S_j$. It has been discussed that the ferroelectricity could be induced by the PM-AFM transition in the centrosymmetric crystals belonging to space groups, like *Pmma*, *Pbnm*, *Pbam*, et.[50] The AFM transition doubles the unit cell and makes such crystals lose their center of symmetry, leading to a non-centrosymmetric point group and inducing the ferroelectricity by the exchange-striction mechanism. In DyVO$_3$, the long-range AFM C-type (or G-type) SO of the V spins is imposed with the centrosymmetric space group *P2$_1$/b* (or *Pbnm*) below $T_{SO1}$, which provides a possibility to lower the symmetries and induce the noncentrosymmetric point group. Furthermore, the real symmetry in most *R*VO$_3$ systems would be lower due to pronounced octahedral tilting.[31,36] Recently, it is argued[31,43] that the symmetry of the G-type OO phase in the *R*VO$_3$ family is lower than the generally accepted *P2$_1$/b*, most likely being non-centrosymmetric monoclinic *Pb*11, as observed by both optical [43,53] and neutron spectroscopy[54] on YVO$_3$ single crystals. The lower symmetry of *Pb*11 allows the possibility of V-V dimerization along *c* axis that is consistent with the tentative identification of an "orbital Peierls state"[43,54] in YVO$_3$ and also accords to the existence of orbital fluctuations in the monoclinic phase in *R*VO$_3$ family[31, 55-57]. It turns out that it is hard[43] to distinguish *Pb*11 and *P2$_1$/b* only from the synchrotron x-ray or neutron diffraction measurements in *R*VO$_3$ system since the monoclinic distortion in these systems is extremely small. Similar optical and neutron spectroscopy experiments performed on YVO$_3$[53-54] are required to distinguish *Pb*11 and *P2$_1$/b* in the monoclinic phase in DyVO$_3$ and other *R*VO$_3$ compounds. All these features support the appearance of ferroelectricity *via* exchange striction below $T_{SO1}$ in DyVO$_3$. In this scenario, the magnetic interactions between adjacent V spin pair modify the V-O-V bond angle/length and induce the oxygen polarization *via* $S_i \bullet S_j$

-type exchange striction, which leads to the macroscopic ferroelectricity in $DyVO_3$. The magnitude of magnetic exchange coupling $J$ [6,51,52] depends on the orbital occupancy and thereby on the shape and orientation of these orbitals. Such scenario is similar to the interpretation of observation of polarization in ortho-manganite $HoMnO_3$[51,58], and spinel $CdV_2O_4$[60], where an AFM ordering of the transition-metal Mn or V spins imposed to the fundamentally centrosymmetric space groups. Thus, the polarization in the temperature region $T_{SO2}<T<T_{SO1}$ results from the cooperative[6,10,61-62] influence of both the exchange striction mechanism (main role) and extended spin-current model, whereas the polarization in the temperature 36 $K < T < T_{SO2}$ should only be ascribed to the exchange striction mechanism.

### F. Origins of the anomalies at 36 K and $T_{SO2}*$ in the ferroelectricity

When the temperature further decreases, an additional weak anomaly is observed only in the $P_b(T)$ curve at 36 K, which is higher than $T_{SO2}^*$. This is in good agreement with the dielectric measurements, *i.e.* a hump for E//b but no anomaly for E//c in the *tanδ* (T) curve at the same temperature of 36 K. Neutron diffraction experiments[42] suggest that, for the warming process, the Dy moments are long-range ordered with $F_xC_y$ mode fully below $T_{SO2}^*$ and then $C_xF_y$ mode below 13 K. However, a common physical picture for the $RVO_3$ family is that the moments of the lanthanide ions are gradually polarized by the ordered V moments via $R$-V coupling at temperatures above the full ordering temperature of the $R^{3+}$ sublattice via $R$-$R$ coupling. For example, in $CeVO_3$, Ce becomes ordered at 28 K, whereas the polarization of the Ce moments starts at about 60 K[63]. In $NdVO_3$, the Nd moments are slightly induced by V moments between 60 and 9.5 K, and then become full ordered at 9.5 K.[43]. For $TbVO_3$, the ordering of the Tb moments with the $C_xF_y$ mode emerges at 11 K, whereas slight ordering of the Tb moments via Tb-V coupling starts at 40

K.[28, 43] A similar polarization process of rare-earth moments has been observed earlier in other perovskites, like NdFeO$_3$[64], and also in iron-pnictide CeFeAsO[65]. Thus, the induced ordering of Dy moments probably takes place at 36 K due to Dy-V coupling. The absence of direct observation on the slight polarization of Dy moment around 36 K in DyVO$_3$ by neutron diffraction may be ascribed to the large neutron absorption cross section of Dy ions, as suggested by Reehuis *et al* in *Ref.* 42. However, they have observed slight ordering of the Tb moments at 40 K from the anomaly in the temperature dependence of magnetic reflections of TbVO$_3$ involving the weaker neutron absorption cross section of Tb ions[43]. Since the Tb-V and the Dy-V coupling are comparable in orthovanadates, it is reasonable that the onset of slight polarization of the lanthanide ions may occur at similar temperatures, *i.e.* at 40 K in TbVO$_3$ and at 36 K in DyVO$_3$. The temperature-induced reentrant transition at $T_{SO2}^*$ and the field-induced switching from G-type SO/C-type OO to C-type SO/G-type OO accompanied by a rapid increase of the Dy moments imply possible exchange coupling between the V *3d* spins and the Dy *4f* moments.[28, 34, 35] Since the long-range Dy ordering ($F_xC_y$ or $C_xF_y$) only takes place in the *ab* plane[42] without component along the *c* axis, the possible partially polarized Dy spins below 36 K would be confined to the *ab* plane. It is well known that ordering of rare-earth moments plays a significant role in inducing the emergence or the anomaly of ferroelectricity, like in HoMnO$_3$[45], TmMn$_2$O$_5$[10], GeFeO$_3$[59] and DyFeO$_3$[66]. Thus, the gradually polarized Dy moments in the *ab* plane by the V moments below $T_{Dy-V}$ (≈36 K) gives rise to an additional ferroelectric distortion via the exchange striction mechanism besides the contribution from the V spin order, which results in a weak increase of $P_b$ with respect to temperature. Since there are no partially polarized Dy spins along the *c* axis, this might account for the absence of an anomaly in the Pc(T) curve around 36 K.

As temperature further decreases to $T_{SO2}^*$ (=18 K), the electric polarization decreases only for E//c, suggesting the concomitant occurrence of a reentrant transition from G-type SO/C-type OO to C-type SO/G-type OO in V sublattice and the emergence of long-range Dy ordering which also influences the ferroelectricity via the exchange striction mechanism. It has come to our attention that very recently, Rajeswaran *et.al*[21] proposed another interesting scenario related to exchange striction mechanism to explain the presence of ferroelectricity below Néel temperature of Cr sublattice in rare-earth orthochromites $R$CrO$_3$ ($R$= Sm, Gd, Tb, Er, Tm) with the centrosymmetric space group *Pbnm*. The poling procedure induces small distortion of $R$ ions and their surrounds, and the interplay between $R$ and Cr providing isotropic Heisenberg-like exchange striction, reinforces such initial distortion, leading to the emergence of ferroelectricity below Néel temperature. This scenario may also be applicable for interpreting the presence of the ferroelectricity below Néel temperature $T_{SO1}$ in our orthovanadate DyVO$_3$. It should be pointed out that the exact microscopic coupling between electric dipoles and spins over a wide temperature range below the Néel temperature $T_{SO1}$ in DyVO$_3$ is unclear due to the complex and numerous competing magnetic exchange interactions in this system. However, we believe the exchange striction mechanism among V spins and/or between Dy and V spins plays a major role in the presence of electric polarization over a wide temperature region with the multiple anomalies with respect to temperature below $T_{SO1}$ in DyVO$_3$. In consideration to the relatively small magnitude of the electric polarization in DyVO$_3$, the combination of high-resolution x-ray techniques performed on orthorhombic $R$MnO$_3$ ($R$=Y[67] or Tb[68]) and the *ab initio* calculation is required to study the ionic displacement and its temperature dependence in order to reveal the exact microscopic origin of the ferroelectricity with multiple anomalies in DyVO$_3$.

## IV. CONCLUSION

In summary, the orthovanadate $DyVO_3$ is found to be a multiferroic material with a high critical temperature of 108 K and very low dielectric loss. The onset of electric polarization along the *b* or *c* axis concurs with the PM-AFM C-type SO transition, where a peak in the heat capacity and an anomaly in the dielectric permittivity are also observed. With decreasing temperature, the $P_b(T)$ curve exhibits two additional anomalies at $T_{SO2}$ (=62 K) and at 36 K, which correspond to transition from C-type SO/G-type OO to G-type SO/C-type OO and the possible onset of polarized Dy moments, respectively. While one anomaly in Pc(T) is also observed at $T_{SO2}$ no anomaly is observed around 36 K and another decrease in Pc(T) is found below $T_{SO2}^*$ (~18 K). The high critical temperature, low dielectric loss and the coupled magnetism and ferroelectricity make $DyVO_3$ a promising insulating multiferroic material. The present study may attract the attention of experimentalists and theorists to investigate the possibility of more high-temperature multiferroics and explore the exact microscopic origin of the ferroelectricity in the $RVO_3$ (*R*=rare earth elements, especially with a relatively small ionic radius such as Gd, Tb, Ho, Er, Yb, Lu) single crystals.

**Figure captions:**

Fig 1. Schematic view of (a) C-type spin ordering (SO) with G-type orbital ordering (OO) and (b) G-type SO with C-type OO of $DyVO_3$. The yellow arrows and blue lobes denote spins and occupied orbitals of the $V^{3+}$ ions, respectively. (c). Summary of the sequence of structural, spin and orbital ordering transitions in $DyVO_3$, based on previous reports[28, 30, 33-37].

Fig. 2(a). Temperature dependence of the field-cooled-cooling (FCC) and field-cooled-warming (FCW) magnetization of $DyVO_3$ in a field of 20 Oe parallel to the b axis, with dashed lines marking the three transitions; (b) Magnetization curve with H//b at 30 K as the field is ramped up and down (see arrows). The inset shows the temperature dependence of the resistivity of $DyVO_3$ in zero magnetic field.

Fig. 3 Temperature dependence of the ratio $C_p$/T for $DyVO_3$ in zero field during the warming process with dashed lines marking the three transitions. The inset shows an enlargement of $C_p$/T around $T^*_{SO2}$.

.

Fig. 4 Temperature dependence of the dielectric permittivity with (a) E// b and (c) E//c for $DyVO_3$, measured at low frequency of 5kHz for warming process. The insets of Figs. 4 (a) and (b) show the corresponding first derivative of the dielectric permittivity as a function of temperature. The dashed lines and the arrow show the locations of the three transitions $T_{SO1}$, $T_{SO2}$ and $T^*_{SO2}$.

Fig. 5 Temperature dependence of the dielectric permittivity and dielectric loss (tan$\delta$) with E// b (a, b) and E//c (c, d) for DyVO$_3$, measured at 100 kHz for both cooling and warming processes. The insets of (b) and (d) show the clear rise of tan$\delta$ at $T_{SO1}$ at a low frequency of 5kHz. The three dashed lines mark the three transitions $T_{SO1}$, $T_{SO2}$, $T^*_{SO2}$ with decreasing temperature.

Fig. 6 Temperature dependence of the dielectric permittivity of DyVO$_3$ in zero magnetic field and in a magnetic field of 5 kOe parallel to the b axis with (a) E// b and (b) E//c.

Fig. 7 (a) Temperature dependence of the electric polarization (P$_b$) of DyVO$_3$, recorded during warming (0.5 K/min) after applying a static electric poling field of +400 or -400 kV/m parallel to the *b* axis; (b) Temperature dependence of the electric polarization (P$_c$) in DyVO$_3$, after using a static electric poling field of +400 kV/m parallel to *c* axis. The inset of Fig. 7(a) shows the corresponding *dP$_b$/dT* curve derived from the measurements after applying a positive poling field.

Phys. Rev. B, 84, 054440 (2011).
68. H. C. Walker et al. Science, 333 1273, (2011).

**Acknowledgements**

This work has been supported by the European project "SOPRANO" under Marie Curie actions (Grant No. PITNGA-2008-214040) and French project "PR Refrigeration Magnétique." We thank F. Guillou for his instructions on the heat capacity measurements and also F. Veillon for the resistivity measurements.

Fig. 1

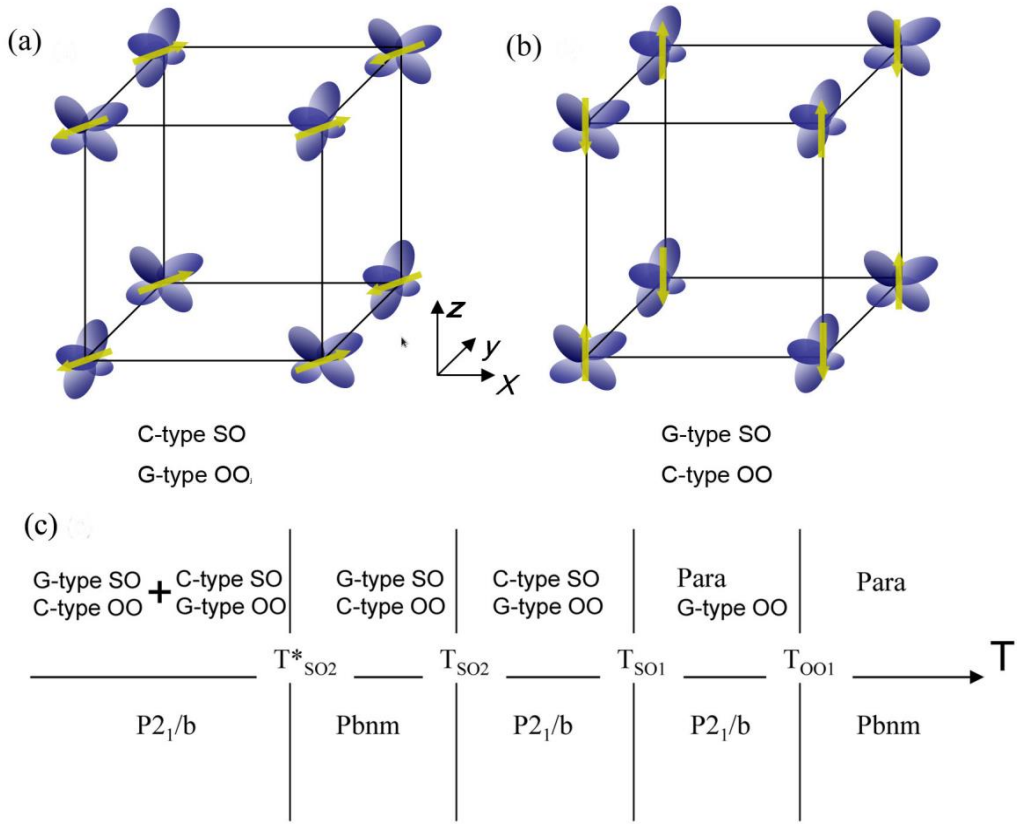

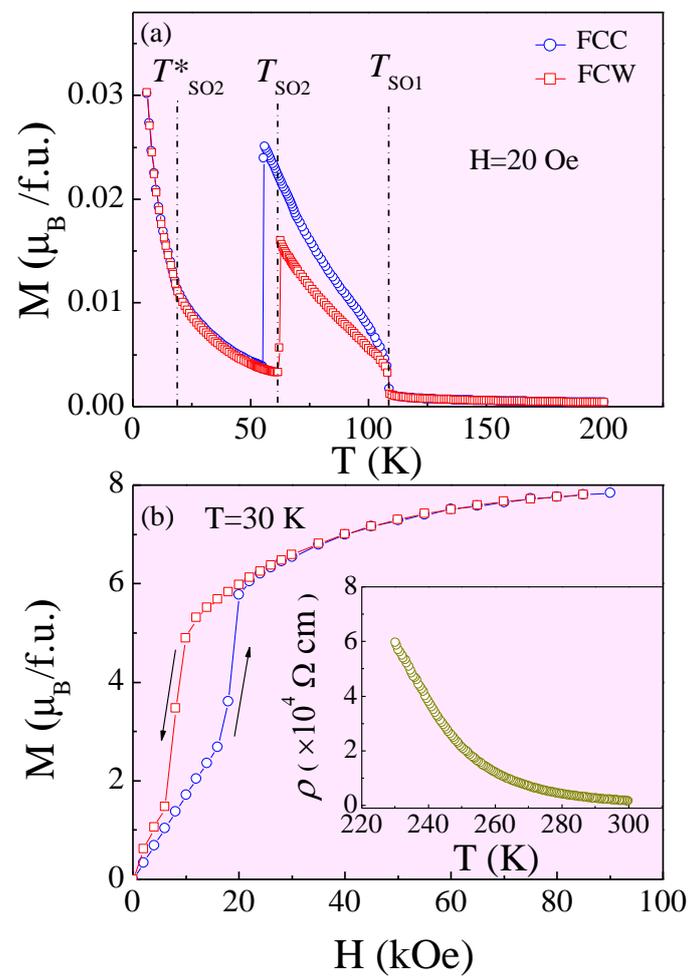

Fig. 3

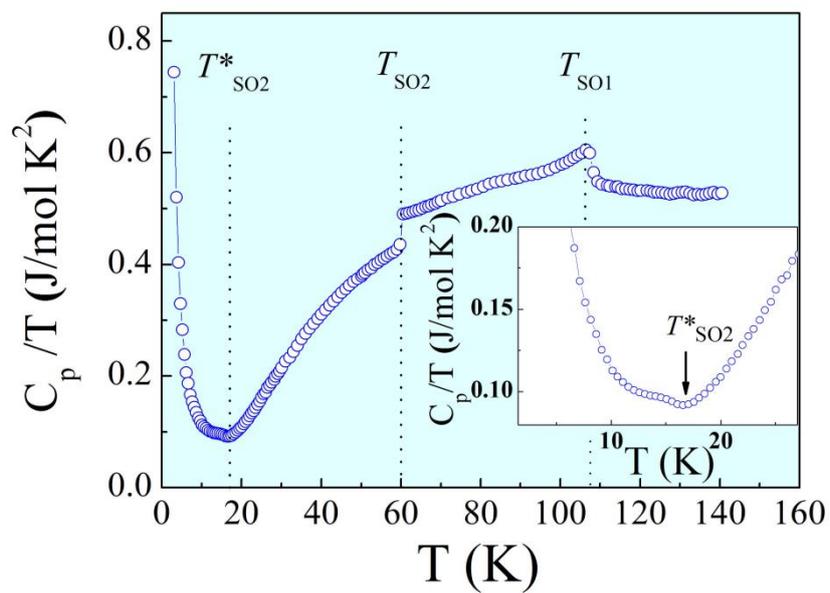

Fig. 4

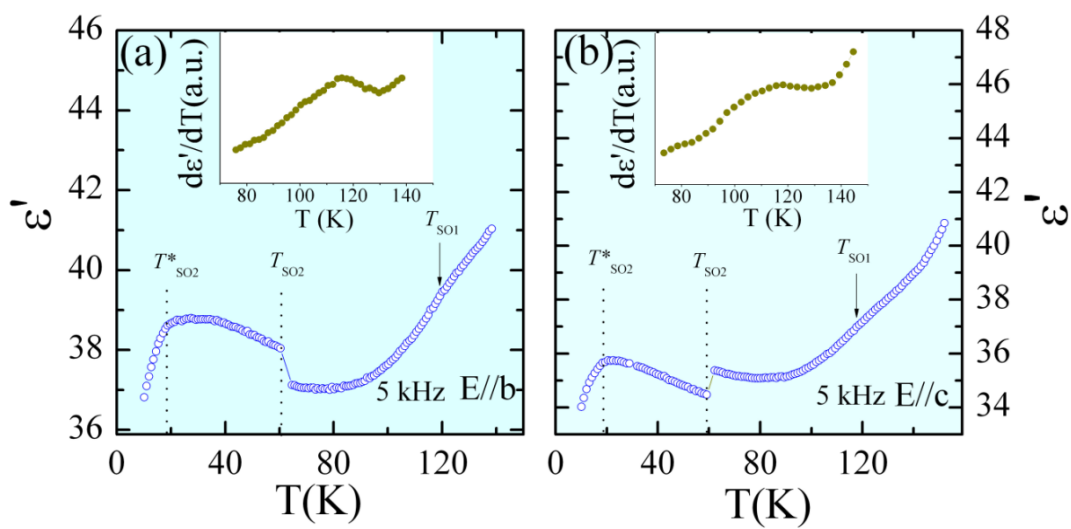

Fig. 5

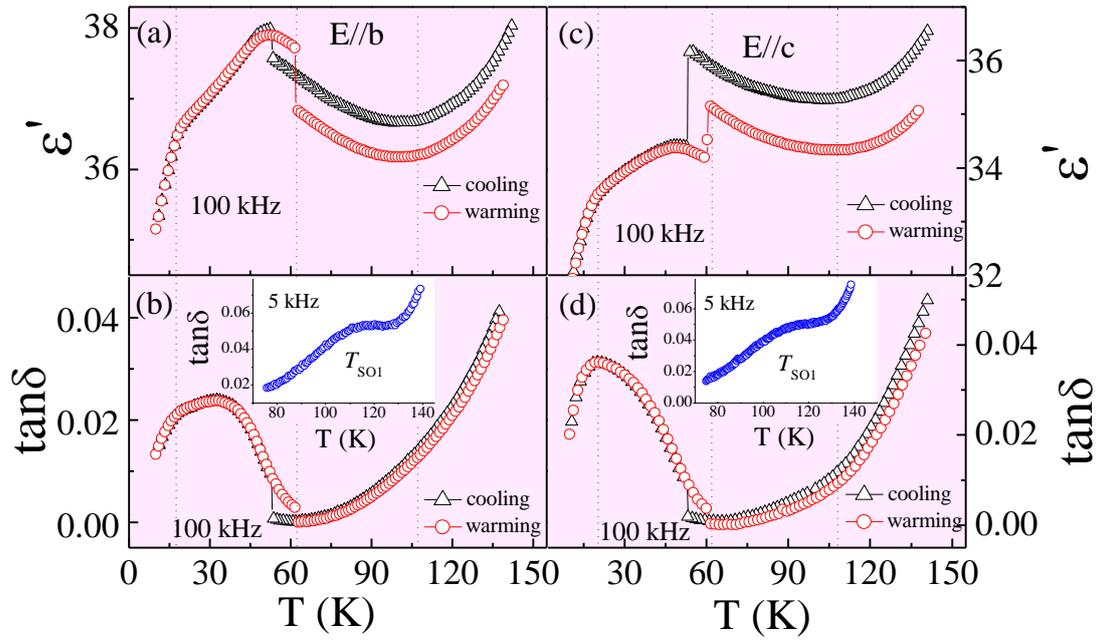

Fig. 6

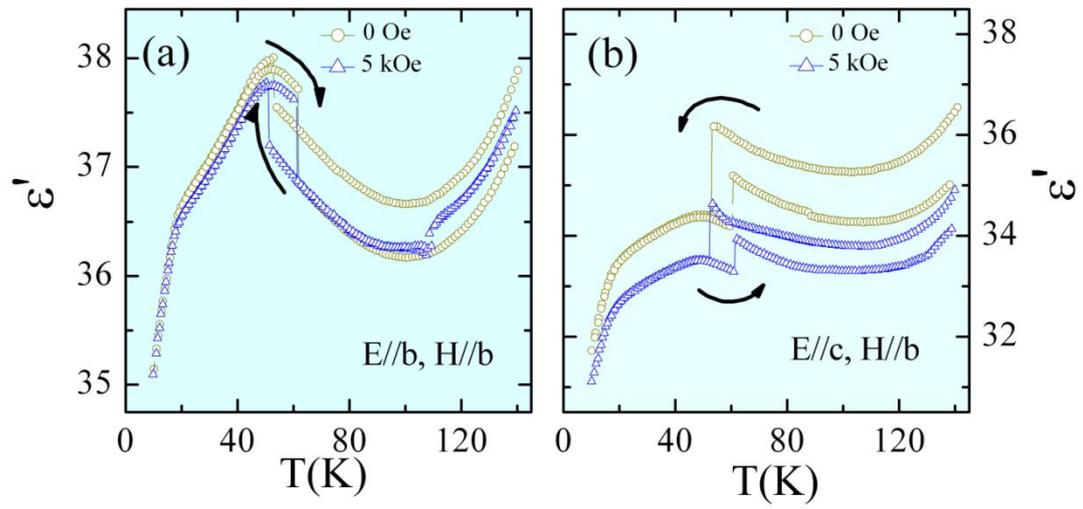

Fig. 7

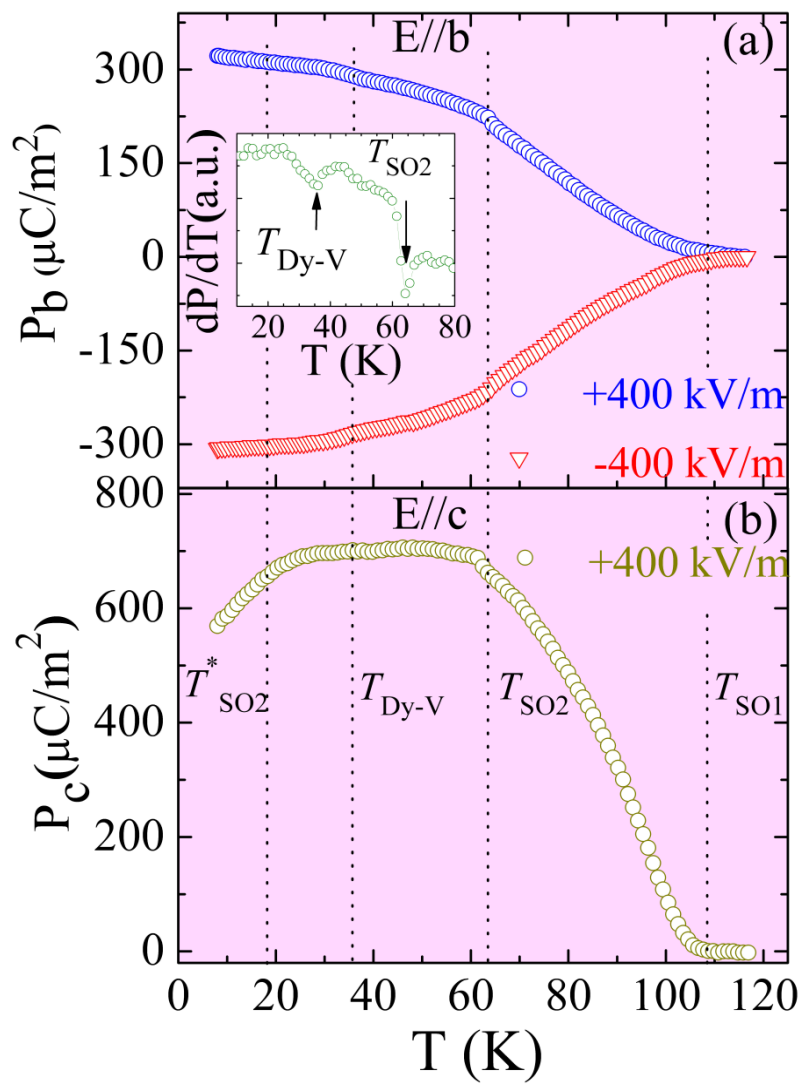

Impact of the various spin and orbital ordering processes on multiferroic properties of orthovanadate $DyVO_3$

Q. Zhang, K. Singh, C. Simon, L. D. Tung, G. Balakrishnan and V. Hardy

After the ferroelectric domains were aligned and a static electric poling field of +400 kV/m was removed at 8 K, we did the short circuit and measured the time dependence of the charge to reach a stable state. The representative results with E//b are shown in Fig. S1. It can be seen that the charge becomes basically stable after waiting time of 5000 seconds. Then, we waited for additional 1400 seconds to further establish the stable charge state. In order to remove the possible contribution of the thermally stimulated current to the pyroelectric current, it is very important to wait for a longer time than relaxation time of stray or trapped charge to remove them (if any) and to reach a stable charge state prior to the P vs T measurements.

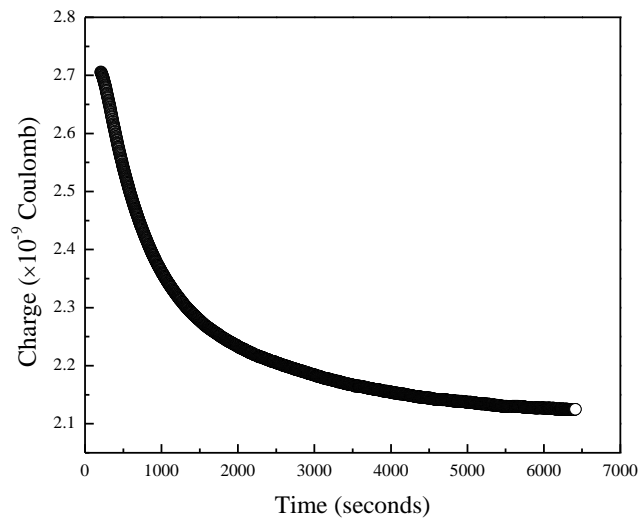

**Fig. S1** Time dependence of the charge at 8 K after a static electric poling field of +400 kV/m with E//b was removed at 8 K